\documentclass[secnumarabic,aps, prd, floatfix,nofootinbib,showkeys,twocolumn,showpacs,10pt,superscriptaddress] 
{revtex4-1}

\usepackage{times}
\usepackage{amsfonts}
\usepackage{amsmath}
\usepackage{amssymb}
\usepackage{natbib}
\usepackage{graphicx}
\usepackage{enumitem}
\usepackage{xcolor}
\usepackage[colorlinks=true,linkcolor=red, citecolor=blue]{hyperref}
\usepackage[outdir=./]{epstopdf}
\usepackage[normalem]{ulem}
\usepackage{amsmath}

\def\araa{ARAA}
\def\aap{A\&A}
\def\aj{AJ}
\def\apj{ApJ}
\def\apjl{ApJL}
\def\apjs{ApJS}

\def\jcap{JCAP}
\def\mnras{MNRAS}
\def\na{New Astronomy}

\begin{document}
\definecolor{orange}{rgb}{0.9,0.45,0}
\def\CovDev{D}
\def\Res{{\mathcal R}}
\def\Gammaflat{\hat \Gamma}
\def\metricflat{\hat \gamma}
\def\Dflat{\hat {\mathcal D}}
\def\part_n{\partial_\perp}
%
\def\Lie{\mathcal{L}}
\def\A{\mathcal{X}}
\def\Aphi{\A_{\phi}}
\def\hAphi{\hat{\A}_{\phi}}
\def\E{\mathcal{E}}
\def\Ham{\mathcal{H}}
\def\M{\mathcal{M}}
\def\R{\mathcal{R}}
\def\p{\partial}
\def\hg{\hat{\gamma}}
\def\hA{\hat{A}}
\def\hD{\hat{D}}
\def\hE{\hat{E}}
\def\hR{\hat{R}}
\def\hcA{\hat{\mathcal{A}}}
\def\hDelt{\hat{\triangle}}
\def\na{\nabla}
\def\dif{{\rm{d}}}
\def\non{\nonumber}
\newcommand{\erf}{\textrm{erf}}
\newcommand{\saeed}[1]{\textcolor{blue}{SF: #1}} 
%
\renewcommand{\t}{\times}
\long\def\symbolfootnote[#1]#2{\begingroup%
\def\thefootnote{\fnsymbol{footnote}}\footnote[#1]{#2}\endgroup}
\title{Matching JWST UV Luminosity Functions with Refined $\Lambda$CDM Halo Models}

\author{Saeed Fakhry}
\email{s\_fakhry@kntu.ac.ir}
\affiliation{Department of Physics, K.N. Toosi University of Technology, P.O. Box 15875-4416, Tehran, Iran}

\author{Maryam Shiravand} 
\email{ma\_shiravand@kntu.ac.ir}
\affiliation{Department of Physics, K.N. Toosi University of Technology, P.O. Box 15875-4416, Tehran, Iran}

\author{Antonino Del Popolo} 
\email{antonino.delpopolo@unict.it}
\affiliation{Dipartimento di Fisica e Astronomia, Universita di Catania, Italy and INFN sezione di Catania, Via S. Sofia 64, I-95123 Catania, Italy}
\date{\today}

\begin{abstract} 
\noindent
The James Webb Space Telescope (JWST) has unveiled a population of unexpectedly massive and luminous galaxies at redshifts $z \gtrsim 7$, posing a significant challenge to the standard $\Lambda$CDM cosmological paradigm. In this work, we address the tension between early JWST observations of luminous high-redshift galaxies and predictions of the standard $\Lambda$CDM model by revisiting the physics of dark matter halo formation. Employing refined halo mass functions derived by Del Popolo {\it el al.} (DP1 and DP2) that incorporate angular momentum, dynamical friction, and redshift-dependent collapse barriers, we demonstrate a significant enhancement in the abundance of massive halos at $z \gtrsim 7$ compared to the conventional Sheth-Tormen (ST) formalism. Using a semi-empirical framework linking halo mass to UV luminosity, we show that the DP2 model reproduces the observed UV luminosity functions from $z=7$ to $14$ with moderate star formation efficiencies, whereas the ST model requires implausibly high efficiencies. Our results suggest that the JWST overabundance problem stems not from new physics beyond $\Lambda$CDM, but from oversimplified treatments of gravitational collapse, highlighting the critical role of small-scale dissipative dynamics in early structure formation.
\end{abstract}


\maketitle
\vspace{0.8cm}

\section{Introduction} 
The Lambda cold dark matter ($\Lambda$CDM) model has been the cornerstone of contemporary cosmology for more than two decades. Its formulation was prompted by the landmark observation of Type Ia supernovae, which revealed that the expansion of the Universe is accelerating \citep{1998AJ....116.1009R}. Since that time, $\Lambda$CDM has offered a robust and consistent framework for understanding a broad spectrum of cosmological phenomena, including the anisotropies in the cosmic microwave background (CMB) and the statistical characteristics of large-scale structure.

To date, direct empirical tests of the standard $\Lambda$CDM cosmological model at redshifts $z \gtrsim 7$ have been profoundly constrained by the limited sensitivity and resolution of traditional cosmological probes. This observational frontier has undergone a paradigm shift with the advent of the James Webb Space Telescope (JWST), whose unprecedented infrared capabilities now permit robust characterization of galaxies during the epoch of reionization and even at earlier cosmic times. Initial JWST observations have uncovered a population of high-redshift galaxy candidates that exhibit stellar masses and ultraviolet (UV) luminosities significantly exceeding, often by factors of several, the predictions of conventional $\Lambda$CDM-based galaxy formation models \citep{2022ApJ...938L..15C, 2022ApJ...940L..14N, 2023Natur.616..266L, 2023ApJS..265....5H, 2023NatAs...7..622C, 2024ApJ...964...71H, 2023NatAs...7..611R}. The presence of such apparently evolved systems within merely $\sim$300-600 million years after the Big Bang presents a substantial challenge to the canonical $\Lambda$CDM framework of hierarchical structure formation, wherein dark matter halos, and the galaxies they host, are expected to assemble progressively through the merging of smaller progenitor structures \citep{2023NatAs...7..731B, 2023MNRAS.518.2511L}.

Reconciling these unexpectedly luminous and massive high-redshift galaxies with the $\Lambda$CDM paradigm has motivated a variety of theoretical and astrophysical explanations. On the astrophysical front, several mechanisms have been proposed to accelerate early stellar mass assembly. For instance, feedback-regulated starbursts in dense, metal-poor environments may enable rapid star formation, thereby enhancing stellar masses well beyond canonical expectations \citep{2023A&A...677L...4P, 2024A&A...689A.244C, 2023MNRAS.523.3201D, 2025ApJ...980..138H}. Moreover, the absence of a fully developed UV background prior to the epoch of reionization could further elevate early star formation rates by reducing photoionization suppression in low-mass halos \citep{2023ApJS..265....5H}. Concurrently, efficient dust ejection in nascent galaxies, driven by supernovae or radiation pressure, may significantly lower dust attenuation, thereby amplifying observed UV luminosities \citep{2023MNRAS.522.3986F, 2024PDU....4401496I}.

While active galactic nuclei (AGNs) have been spectroscopically confirmed in a handful of high-redshift systems, most notably CEERS\_1019 and GN-z11 \citep{2023A&A...677A..88B, 2023ApJ...952...74T, 2023ApJ...953L..29L, 2023ApJ...959...39H}, their scarcity implies that stellar processes dominate the integrated light of the majority of JWST-detected high-$z$ galaxies. An alternative proposition posits a top-heavy initial mass function (IMF), skewed toward massive stars, which would enhance UV emissivity per unit stellar mass \citep{2023ApJ...951L..40S, 2024MNRAS.534..523C}. However, this scenario remains tightly constrained by the concomitant effects of strong stellar feedback, which can suppress further star formation, and by the current lack of direct observational evidence supporting a systematically top-heavy IMF at these epochs \citep{2024A&A...686A.138C}. Concurrently, numerous cosmological alternatives have been explored, including dynamical dark energy \citep{2023JCAP...10..072A, 2024JCAP...07..072M}, modifications to the primordial power spectrum \citep{2023MNRAS.526L..63P, 2023PhRvD.107d3502H, 2024PhLB..85839062B}, primordial black hole dark matter \citep{2023arXiv230617577H, 2024SCPMA..6709512Y}, early dark energy \citep{2024MNRAS.533.3923S, 2024JCAP...05..097F, 2025PhRvD.111b3519J}, non-standard dark matter models \citep{2024MNRAS.534.2848D}, and theories of modified gravity \citep{2024arXiv241203534M, 2025arXiv250111103S}.

Among the most striking revelations from JWST observations of the high-redshift galaxies is the unexpected behavior of the UV luminosity function (UVLF) \citep{2023MNRAS.518.6011D, 2025asi..confP.121P}. The UVLF represents the comoving number density of galaxies per unit rest-frame UV magnitude and constitutes a cornerstone diagnostic for probing galaxy formation and star formation activity in the early universe. At high redshifts, UV emission is predominantly generated by short-lived, massive stars, rendering the UVLF a near-instantaneous tracer of recent star formation \citep{2025Natur.639..897W, 2025MNRAS.540.2176C}. Of particular importance is the bright end of the UVLF, which is exquisitely sensitive to the rarest, most massive dark matter halos, precisely those now being resolved by the JWST \citep{2024ApJ...968...79L}. Recent JWST observations have uncovered a surplus of luminous galaxies at high-redshifts relative to the predictions of pre-JWST theoretical models, thereby challenging prevailing assumptions about the efficiency of star formation and the efficacy of feedback mechanisms within the most massive halos during cosmic dawn \citep{2024Natur.633..318C, 2025PhRvD.111b3519J}. Empirically, the UVLF at these epochs is often well described by a double power-law form, whose shape encodes the interplay between the underlying dark matter halo mass function and baryonic physics \citep{2023MNRAS.518.6011D}.

On the other hand, a suitable dark matter halo model serves as a cornerstone in determining the UVLF at high redshifts \citep{2020MNRAS.495.3602W, 2024OJAp....7E.105O, 2025MNRAS.536..988F}. A model that incorporates effective physical and geometric factors to more precisely characterize the conditions for the direct collapse of density fluctuations will inherently offer a more accurate representation of the corresponding cosmic phenomenon, see, e.g., \citep{2021PhRvD.103l3014F, 2022PhRvD.105d3525F, 2022ApJ...941...36F, 2023PDU....4101244F, 2023PhRvD.107f3507F, 2023ApJ...947...46F, 2023arXiv230811049F, 2024PDU....4601544F, 2024arXiv240115171F, 2024PDU....4601712T, 2025ApJ...989..116F}. In the $\Lambda$CDM paradigm, galaxies reside within dark matter halos, and their rest-frame UV luminosities are fundamentally linked to halo mass through the complex interplay of baryonic physics, including gas accretion, radiative cooling, star formation efficiency, and feedback processes. This connection enables the UVLF to be constructed by establishing a luminosity-halo mass relation, typically calibrated via abundance matching, empirical modeling, or semi-analytic frameworks anchored to cosmological simulations \citep{2020MNRAS.495.3602W}. Critically, the bright end of the UVLF is dictated by the high-mass exponential decline of the halo mass function (HMF), rendering it exquisitely sensitive to the number density of the rarest, most massive halos. Consequently, robust theoretical forecasts of the UVLF, particularly at $z \gtrsim 7$, where JWST is now probing, demand a highly accurate HMF, as uncertainties in halo collapse dynamics, non-linear structure growth, or potential departures from standard cosmological assumptions can profoundly alter inferences about early galaxy formation and evolution.

For over two decades, the Sheth-Tormen (ST) HMF has been a cornerstone for modeling HMF in cosmological studies, particularly through $N$-body simulations \citep{1999MNRAS.308..119S, 2001MNRAS.323....1S}. Derived from an ellipsoidal collapse framework, it refines the original Press-Schechter (PS) theory by better approximating the geometry of overdense region collapse. Despite its successes, the ST model remains phenomenological, lacking detailed physical mechanisms such as tidal torques, dynamical friction, or the influence of the cosmological constant on collapse thresholds, which become particularly significant at high redshifts \citep{2006ApJ...637...12D, 2017JCAP...03..032D}. This limitation, though not a failure of the $\Lambda$CDM paradigm itself, results in almost inaccurate predictions for halo abundances in high-$\sigma$ peaks that give rise to the first galaxies. To address these shortcomings, we turn to the more physically comprehensive models of \citep{2006ApJ...637...12D, 2017JCAP...03..032D}, which incorporate angular momentum, dynamical friction, and cosmological constant effects into the collapse dynamics. These models can offer mass- and redshift-dependent barriers, potentially enhancing the abundance of massive halos at early epochs, critical for explaining the luminous galaxies observed by JWST, e.g., \citep{2025arXiv250723742F}.

In this work, we investigate the UV luminosity function of galaxy candidates identified at redshifts $z \gtrsim 7$ with JWST, within the framework of refined dark matter halo models. The structure of the paper is as follows. In Section \ref{sec:ii}, we present an analytical treatment of the matter power spectrum in the standard cosmological model. In Section \ref{sec:iii}, we introduce improved halo mass functions that incorporate both geometric and physical modifications. In Section \ref{sec:iv}, we construct a theoretical framework for the UV luminosity function and provide a comparison between the relevant predictions of these halo models and JWST observations of high-redshift galaxy candidates. Finally, in Section \ref{sec:v}, we summarize our findings and consider their broader implications.

\section{Matter Power Spectrum in the \(\Lambda\)CDM Model}\label{sec:ii} 

The matter power spectrum serves as a fundamental descriptor of the statistical distribution of matter density fluctuations across different spatial scales in the Universe. Within the $\Lambda$CDM framework, it connects primordial perturbations generated during inflation to the present-day large-scale structure via gravitational growth.

The matter overdensity field $\delta(\mathbf{x})$ is defined as the fractional deviation of the local matter density $\rho(\mathbf{x})$ from the mean density $\bar{\rho}$:
\begin{equation}
\delta(\mathbf{x}) = \frac{\rho(\mathbf{x}) - \bar{\rho}}{\bar{\rho}}.
\end{equation}
Transforming to Fourier space, the overdensity becomes $\tilde{\delta}(\mathbf{k})$, and the matter power spectrum $P(k)$ is defined through the two-point correlation function in Fourier domain as
\begin{equation}
\langle \tilde{\delta}(\mathbf{k}) \tilde{\delta}^*(\mathbf{k}') \rangle = (2\pi)^3 \delta_D(\mathbf{k} - \mathbf{k}') P(k),
\end{equation}
where $\delta_D$ is the Dirac delta function, enforcing statistical isotropy and homogeneity.

To facilitate an intuitive understanding, the dimensionless power spectrum $\Delta^2(k)$, which describes the contribution to variance per logarithmic interval in $k$, is commonly used:
\begin{equation}
\Delta^2(k) = \frac{k^3}{2\pi^2} P(k).
\end{equation}

In the linear regime, the $\Lambda$CDM matter power spectrum at redshift $z$ can be expressed as
\begin{equation}
P(k, z) = A_s k^{n_s} T^2(k) D^2(z),
\end{equation}
where $A_s$ is the primordial scalar amplitude determined by inflation, $n_s$ is the scalar spectral index (with measured value $n_s \approx 0.96$). The transfer function, $T(k)$, encapsulates the physical processes that shape perturbations from horizon entry through matter domination. It can be computed using numerical Boltzmann solvers or approximated with analytic fitting formulas, such as those proposed by Eisenstein and Hu \citep{1998ApJ...496..605E}, which account for the effects of baryon acoustic oscillations and matter–radiation equality. Also, $D(z)$ is the linear growth factor describing the amplitude evolution of density fluctuations with redshift \citep{1998ApJ...496..605E}. In a flat $\Lambda$CDM Universe, $D(z)$ can be determined by \footnote{Note that the normalization condition $D(0) = 1$ is adopted.}
\begin{equation}
D(z) = \frac{5 \Omega_{m,0} H_0^2}{2} H(z) \int_z^\infty \frac{1+z'}{H^3(z')} \, dz',
\end{equation}
where $\Omega_{m,0}$ is the present matter density parameter, $H_0$ is the Hubble constant, and $H(z)$ is the Hubble parameter evolving with redshift.

The characteristic turnover in the power spectrum marks the comoving horizon scale at matter-radiation equality, while baryon acoustic oscillations imprint periodic features that serve as robust standard rulers in cosmology. Current observational constraints from galaxy redshift surveys, Lyman-$\alpha$ forest data, and measurements of the CMB anisotropies tightly agree with the $\Lambda$CDM matter power spectrum, enabling precision cosmology and detailed understanding of structure formation \citep{2022ApJ...928L..20S, 2019MNRAS.489.2247C}.
\section{Halo Mass Functions}\label{sec:iii} 
The halo mass function can be precisely formulated using excursion set theory, which considers the density field as a stochastic process spanning various scales. This approach is fundamentally based on the spherical collapse model, which determines the critical overdensity needed for the gravitational collapse of spherically symmetric perturbations \citep{1974ApJ...187..425P}. In an Einstein-de Sitter Universe, assuming ideal spherical symmetry, this critical overdensity threshold is given analytically by: 
\begin{equation}
\delta_{\rm sc}=\frac{3(12\pi)^{2/3}}{20}\left(1-0.01231\log\left[1+\frac{\Omega_{\rm m}^{-1}-1}{(1+z)^{3}}\right]\right),
\end{equation}
where $\Omega_{\rm m}$ represents the matter density parameter. The critical overdensity threshold is commonly approximated as $1.686$ within a restricted range of redshifts. Nevertheless, a significant drawback of this traditional method is that $\delta_{\rm sc}$ shows only a weak dependence on mass, which can result in systematic underestimations in specific mass ranges.

The initial improvement to the PS threshold was introduced by \citep{1998A&A...337...96D}, showing that the collapse threshold can be modified to include mass-dependent corrections expressed as:
\begin{equation}
\delta_{\rm cm}=\delta_{\rm sc}\left(1+\frac{\beta}{\nu^{\alpha}}\right),
\end{equation}
using best-fit parameter values of $\alpha=0.585$ and $\beta=0.46$. In \citep{2001MNRAS.323....1S}, an alternative collapse threshold was developed by taking into account ellipsoidal, rather than spherical, collapse geometry:
\begin{equation}
\delta_{\rm ec}=\delta_{\rm sc}\left(1+\frac{\beta_1}{\nu^{\alpha_1}}\right),
\end{equation}
where $\alpha_1 = 0.615$,  $\beta_1 = 0.485$. Also, $\nu$ denotes the peak height, which is precisely defined as:
\begin{equation}
\nu \equiv \frac{\delta_{\rm sc}(z)}{\sigma_M}=\frac{1.686}{D(z)\sigma_M}.
\end{equation}
In this equation, $\sigma_M$ represents the linear root-mean-square amplitude of the density fluctuations at comoving scales associated with the halo mass $M$:
\begin{equation} 
\sigma^2_{M} = \frac{1}{2\pi^{2}}\int_{0}^{\infty} P(k) W^2(kR) k^{2} {\rm d}k,
\end{equation}
where $W(kR)$ is the filtering window function applied at the comoving scale $R$. The halo mass relates to the comoving scale through $M=4\pi \bar{\rho}_{\rm m,0} R^3/3$, where $\bar{\rho}_{\rm m,0}$ represents the current mean comoving density of dark matter. In this study, we have used a real-space top-hat window function defined by $W(x)=3(\sin x - x\cos x)/x^3$.

The excursion set framework provides the basis for determining the unconditional mass function of dark matter halos, which measures the average comoving number density of halos within each logarithmic mass interval. This key relation is given by:
\begin{equation} 
n(M, z) = \frac{\bar{\rho}_{\rm m,0}}{M^2} \left|\frac{{\rm d}\log\nu}{{\rm d}\log M} \right| \nu f(\nu),
\end{equation}
where $\nu f(\nu)$ represents the multiplicity function, which describes the distribution of the first crossing. The original PS approach provides the following analytical form for this multiplicity function \citep{1974ApJ...187..425P}:
\begin{equation}
[\nu f(\nu)]_{\rm PS}=\sqrt{\frac{2}{\pi}}\frac{(\nu+0.556)\exp[-0.5(1+\nu^{1.34})^{2}]}{(1+0.0225\nu^{-2})^{0.15}}.
\end{equation}

\begin{figure*}
\centering
\includegraphics[width=0.9\linewidth]{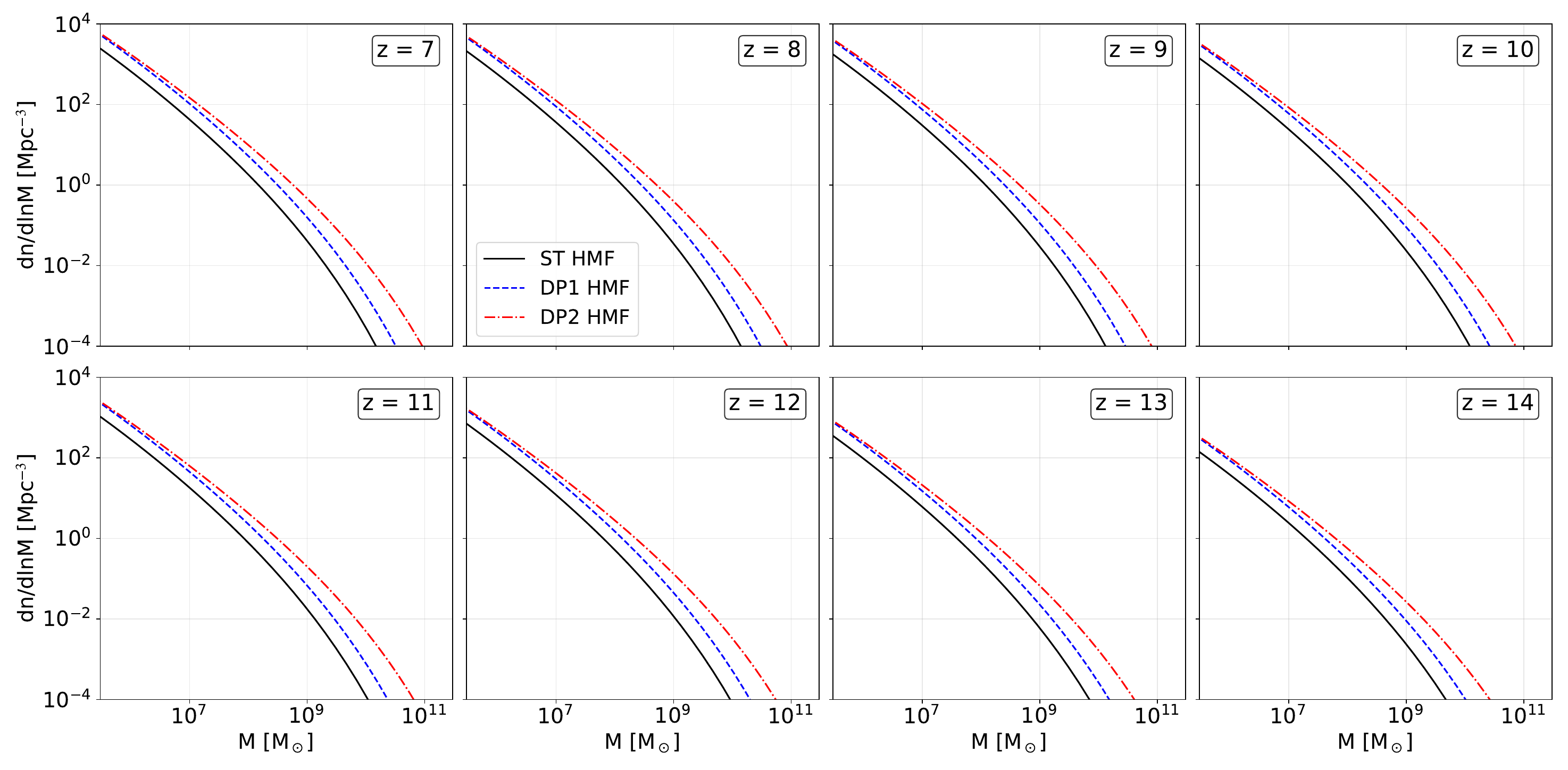}
\caption{ST, DP1 and DP2 halo mass functions at redshifts $z=7\mbox{-}14$.}
\label{Fig1}
\end{figure*}

Despite its foundational role, the PS formalism shows significant deviations from the dark matter halo distributions obtained in high-resolution numerical simulations. These differences stem from physical effects that the original PS model does not fully capture, though they play a crucial role in shaping halo abundance statistics. To overcome these shortcomings, the ST mass function introduces geometric adjustments and generalizes the spherical collapse model to ellipsoidal collapse scenarios. The ST mass function is expressed analytically as \citep{2001MNRAS.323....1S}:
\begin{equation}
[\nu f(\nu)]_{\rm ST}=A_{1}\sqrt{\frac{2\nu^{\prime}}{\pi}}\left(1+\frac{1}{\nu^{\prime q}}\right)\exp\left(-\frac{\nu^{\prime}}{2}\right),
\end{equation}
In the above relation, the parameters are set as $q=0.3$, $\nu^\prime=0.707\nu^2$, and $A_{1}=0.322$, where the normalization constant $A_1$ is fixed by enforcing the condition $\int f(\nu) d\nu = 1$.

In addition to geometric factors involved in halo virialization, several other physical processes play a crucial role in the dynamics of overdensity collapse and, consequently, influence the halo mass function. Accounting for these effects is vital as they capture the essential physics governing halo formation and evolution, as well as the processes underpinning hierarchical structure growth throughout cosmic time. This framework allows the collapse threshold to vary according to physically motivated parameters, enabling the barrier to respond dynamically and providing more precise theoretical predictions for halo collapse. Key modifications include the influence of angular momentum, dynamical friction, and the cosmological constant \footnote{While the cosmological constant has a negligible impact at high redshifts, it is included here for theoretical completeness}. These enhancements reduce systematic discrepancies, especially in mass ranges where observational data remain uncertain.

By thoroughly incorporating the effects of angular momentum and the cosmological constant, the resulting mass function, referred to as DP1 in this study, offers improved theoretical precision \citep{2006ApJ...637...12D}:
\begin{equation}
[\nu f(\nu)]_{\rm DP1}=A_{2}\sqrt{\frac{\nu^{\prime}}{2\pi}}k(\nu^{\prime})\exp\left[-0.4019\nu^{\prime} l(\nu^{\prime})\right],
\end{equation}
here, $A_{2} = 0.974$ denotes the normalization constant, and:
\begin{equation}
k(\nu^\prime)=\left(1+\frac{0.1218}{\nu^{\prime 0.585}}+\frac{0.0079}{\nu^{\prime 0.4}}\right),
\end{equation}
\begin{equation}
l(\nu^\prime)=\left(1+\frac{0.5526}{\nu^{\prime 0.585}}+\frac{0.02}{\nu^{\prime 0.4}}\right)^2.
\end{equation}

The impact of dynamical friction on the collapse barrier was thoroughly examined by \citep{2017JCAP...03..032D}, resulting in the mass function labeled DP2:
\begin{equation}
[\nu f(\nu)]_{\rm DP2}=A_{3}\sqrt{\frac{\nu^\prime}{2\pi}}m(\nu^\prime)\exp[-0.305\nu^{\prime2.12} n(\nu^\prime)],
\end{equation}
where in this case, $A_{3} = 0.937$ serves as the normalization factor, and:
\begin{equation}
m(\nu^\prime)=\left(1+\frac{0.1218}{\nu^{\prime 0.585}}+\frac{0.0079}{\nu^{\prime 0.4}}+\frac{0.1}{\nu^{\prime 0.45}}\right),
\end{equation}
\begin{equation}\label{nnuprime}
n(\nu^\prime)=\left(1+\frac{0.5526}{\nu^{\prime 0.585}}+\frac{0.02}{\nu^{\prime 0.4}}+\frac{0.07}{\nu^{\prime 0.45}}\right)^2.
\end{equation}

In Fig.\,\ref{Fig1}, we have presented a comparative analysis of three distinct HMFs: the standard ST model and the two physically motivated alternatives, DP1 and DP2 across the high-redshift range $7\leq z \leq 14$. Each panel shows the comoving number density of dark matter halos per logarithmic mass interval, ${\rm d}n/{\rm d}\ln M$, as a function of halo mass $M$. A striking feature evident across all redshifts is that the DP1 and especially DP2 models predict an enhanced abundance of massive halos $M \gtrsim 10^{8} \, M_\odot$ compared to the ST model. 

At lower redshifts such as $z=7\sim 8$, the ST model still yields a non-negligible population of halos up to $\sim 10^{10} \, M_\odot$, but it falls short in the high-mass tail relative to DP1 and DP2. The DP1 model, which incorporates angular momentum effect into the collapse barrier, already shows a modest boost in massive halo abundance. However, the DP2 model, further augmented by the inclusion of dynamical friction, demonstrates a dramatic increase in the number density of halos above $10^{9} \, M_\odot$. This is a direct consequence of the lowered effective collapse threshold for massive peaks, allowing more regions to collapse earlier than predicted by ellipsoidal-collapse-based models like ST.

As redshift increases beyond  $z=10$, the differences between the models become stark. By $z=12\sim 14$, the ST mass function essentially truncates above $\sim 10^{10} \, M_\odot$, predicting vanishingly few halos capable of hosting the luminous galaxies recently detected by JWST. In contrast, both DP1 and DP2 continue to produce a non-zero population of halos up to $\sim 10^{10} \, M_\odot$, with DP2 consistently outperforming DP1 in the high-mass regime. This behavior highlights the critical role of dissipative physics, particularly dynamical friction, in enabling early structure formation. The exponential sensitivity of the high-mass tail of the HMF to the collapse threshold means that even small physical corrections, negligible at low redshift, become dominant at cosmic dawn.

The redshift evolution shown in Figure 1 also reveals a systematic trend that is not merely a numerical artifact but reflects the underlying physics of hierarchical structure formation in a $\Lambda$CDM universe with a finite linear growth factor $D(z)$. At higher $z$, $D(z)$ is smaller, so only peaks with exceptionally high initial overdensities can collapse. The ST model, with its fixed ellipsoidal barrier, underestimates how often such peaks occur. In contrast, DP1 and DP2 feature mass- and redshift-dependent collapse thresholds that respond to physical effects like tidal torques and dynamical friction, thereby increasing the predicted abundance of rare, massive halos precisely when they are most observationally relevant.
\section{UV Luminosity Function and JWST Observations} \label{sec:iv}
The rest-frame UV luminosity function (UVLF) is a key probe of galaxy formation during the epoch of reionization, as UV photons at $\lambda \sim 1500$\,\AA\ are predominantly produced by massive, short-lived stars and thus directly trace recent star formation \citep{2014ARA&A..52..415M, 2015ApJ...813...21M, 2023MNRAS.521..497M}. Measurements of the UVLF at $z \gtrsim 7$ therefore provide a window into both the growth of galaxies and the ionizing photon budget available to reionize the intergalactic medium \citep{2015ApJ...803...34B, 2015ApJ...810...71F}. 

Following the semi-empirical approach of \citep{2015ApJ...813...21M}, one can link star formation in galaxies to the growth of their host dark matter halos. In this framework, the star formation rate (SFR) of a halo of mass $M_h$ is approximated as 
\begin{equation} \label{eqeq1}
SFR(M_{h})=f_b \, f_{\star}(M_h, z)\,\frac{M_{h}}{t_{\rm sf}}, 
\end{equation} 
where $f_b = \Omega_b/\Omega_m$ is the cosmic baryon fraction, $f_\star(M_h, z)$ is the star formation efficiency that generally depends on halo mass and redshift \citep{2015ApJ...813...21M, 2016MNRAS.460..417S, 2023ApJ...955L..35S}, and $t_{\rm sf}$ is a characteristic star formation timescale. This prescription assumes that baryons accrete into halos at the same rate as dark matter and are converted into stars with an efficiency that depends primarily on halo mass \citep{2010ApJ...714L.202T, 2013ApJ...768L..37T}. The SFR is then converted to UV luminosity using the calibration of \citep{2014ARA&A..52..415M}:
\begin{equation}\label{eqeq2}
L_\nu =\frac{SFR}{K_{\rm UV}} ,
\end{equation} 
where $K_{\rm UV} \simeq \mathcal{O}(10^{-28})$ ${\rm M_\odot \, yr^{-1}} / ({\rm erg \, s^{-1} \, Hz^{-1}})$. The corresponding rest-frame UV magnitude is
\begin{equation}\label{eqeq3}
M_{\rm UV} = -2.5 \log_{10} L_\nu + 51.63 .
\end{equation}
The UVLF, $\Phi(M_{\rm UV})$, is derived by mapping the HMF to magnitudes \citep{2015ApJ...813...21M, 2023MNRAS.521..497M}:
\begin{equation}
\Phi(M_{\rm UV}) = \frac{dn}{dM_h} \left| \frac{dM_h}{dM_{\rm UV}} \right|.
\end{equation}
For the HMF, we adopt the ST, DP1, and DP2 formalisms and compute $\Phi(M_{\rm UV})$ numerically by first evaluating the HMF using analytic prescriptions, then converting halo mass to SFR using Eq.\,(\ref{eqeq1}), mapping the SFR to UV magnitude via Eqs.\,(\ref{eqeq2}) and (\ref{eqeq3}), and finally applying the Jacobian $\left| dM_h/dM_{\rm UV}\right|$ to transform from halo mass to magnitude space. In this study, we adopt two distinct approaches to modeling star formation efficiency. First, we treat it as a free parameter, allowed by physically permissible values relative to halo mass $10^{10}\lesssim M_h/M_{\odot}\lesssim 10^{12} $, followed by the findings reported in \citep{2015ApJ...813...21M, 2016MNRAS.460..417S}. Second, we model it as a function of both halo mass and redshift (see Eq.\,(9) of \citep{2016MNRAS.460..417S}).

\begin{figure*}
\centering
\includegraphics[width=0.9\linewidth]{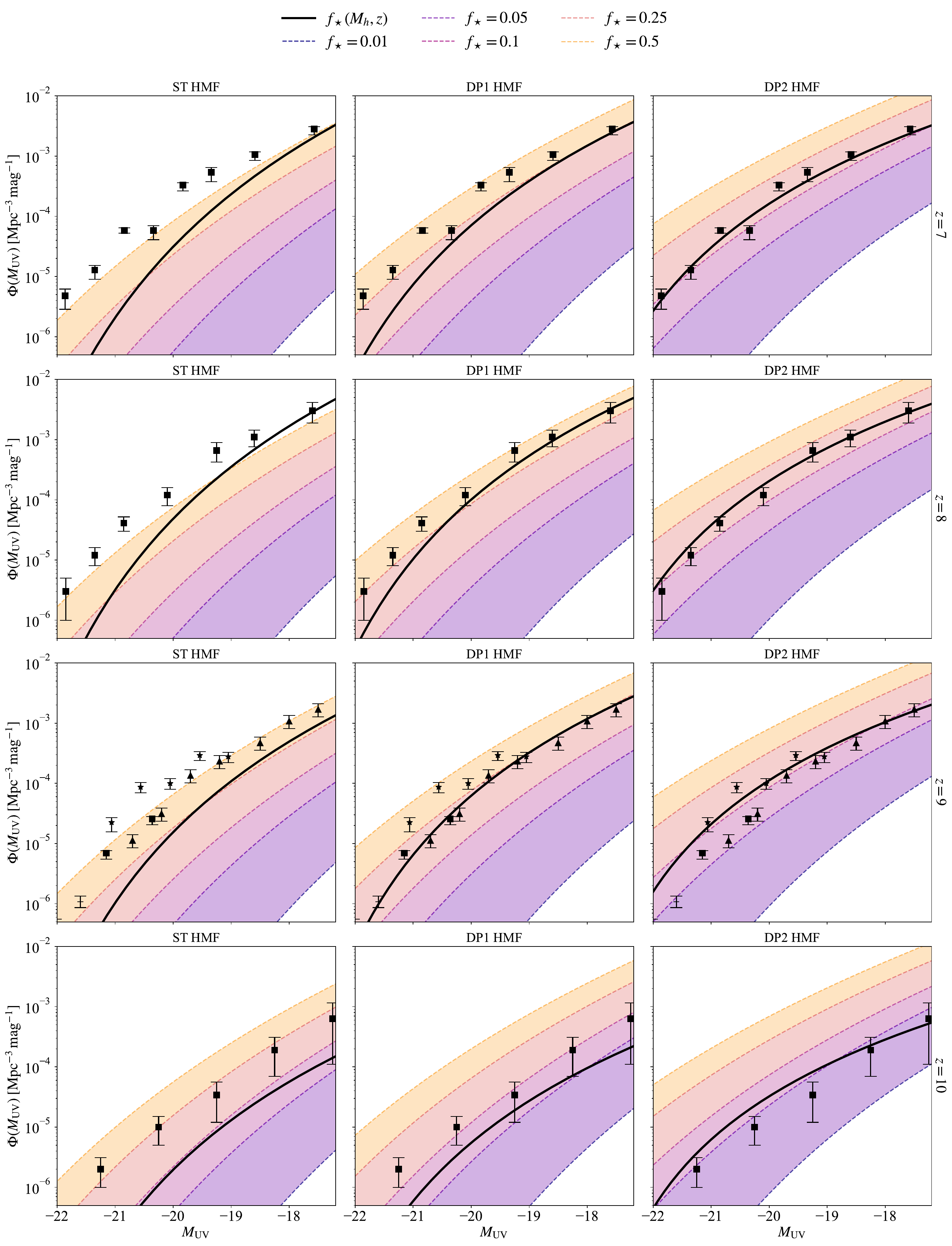}
\caption{The UVLF as a function of UV magnitude for different star formation efficiencies, evaluated using the ST (left panel), DP1 (middle panel), and DP2 (right panel) mass functions at redshifts $z=7\mbox{-}10$. The black lines represent the UVLFs predicted by the different models incorporating the redshift- and halo-mass-dependent star formation efficiency. The dataset with error bars represent recent observational constraints on the UV luminosity function at the same redshift from multiple studies: filled squares denote measurements from \citet{2015ApJ...803...34B, 2021AJ....162...47B, 2022ApJ...927...81B}; filled stars correspond to \citet{2023ApJ...946L..13F}; upward filled triangles represent \citet{2023MNRAS.520.4554D}; and dash indicate \citet{2021ApJ...922...29S}.}
\label{Fig2}
\end{figure*}

In Fig.\,\ref{Fig2}, we have depicted the UVLF as a function of rest-frame UV magnitude across four redshifts $z = 7, 8, 9$, and $10$ and three HMFs: the standard ST model (left column), and the two physically motivated alternatives DP1 and DP2 (middle and right columns, respectively). Each panel displays theoretical predictions for five star formation efficiencies $f_\star = 0.01, 0.05, 0.1, 0.25$, and $0.5$. Additionally, the corresponding results incorporating $f_\star(M_h, z)$ are presented in each panel as solid black lines. We have also included a dataset with error bars, which reflects recent observational constraints on the UV luminosity function at a common redshift from multiple studies: filled squares correspond to the measurements presented in \citet{2015ApJ...803...34B, 2021AJ....162...47B, 2022ApJ...927...81B}; filled stars denote the results from \citet{2023ApJ...946L..13F}; upward-pointing filled triangles represent the data reported by \citet{2023MNRAS.520.4554D}; and dashed lines indicate the constraints derived by \citet{2021ApJ...922...29S}. The UVLF is expressed in comoving number density per unit magnitude per cubic megaparsec and spans the bright-end regime $(M_{\rm UV} \lesssim -17)$ where JWST constraints are most robust.

At $z = 7$, the ST model (left panel) only approaches the observational data for the highest allowed star formation efficiency $(f_\star \gtrsim 0.5)$, and even then underpredicts the UVLF of galaxies in several bins of UV magnitudes. For moderate efficiencies $(f_\star \leq 0.25)$, the ST predictions fall significantly below the data, by more than an order of magnitude at the bright end, highlighting a clear tension with JWST observations unless extreme astrophysical assumptions are invoked. In contrast, the DP1 model (middle panel) shows markedly improved agreement: even at $f_\star = 0.25$, the predicted UV LF aligns with observations within uncertainties across most of the magnitude range, and at $f_\star \lesssim 0.5$ it matches the bright-end data almost perfectly. The DP2 model (right panel) performs even better, achieving excellent agreement with the data for $f_\star \simeq 0.15$. This demonstrates that incorporating angular momentum, dynamical friction, and cosmological constant effects into the collapse barrier substantially enhances the predicted abundance of luminous high-$z$ galaxies. The corresponding predictions that incorporate $f_\star(M_h, z)$ exhibit a nearly identical trend. Specifically, the UVLF derived from the ST model systematically underpredicts the observational data. Moreover, the underprediction exhibited by the ST model becomes increasingly pronounced with rising galaxy luminosity. The DP1 model improves upon this, yielding results that are somewhat closer to the observations. Finally, the UVLF derived from the DP2 model demonstrates good agreement with the observational data.

At $z = 8$, the discrepancy for the ST model becomes a bit more pronounced. Even with high star formation efficiencies, the ST predictions fall below the observed UVLF at all UV magnitudes, and for $f_\star \leq 0.25$, the model underpredicts by up to an order of magnitude. The DP1 model again shows significant improvement: with $0.25 \lesssim f_\star \lesssim 0.5$, it closely tracks the observational data across the full magnitude range shown. The DP2 model continues to outperform both, matching the bright-end data ($M_{\rm UV} \lesssim -20$) even at $f_\star = 0.1$, and providing a near-perfect fit at $f_\star \simeq 0.25$. Specifically, when the parameter $f_\star(M_h, z)$ is taken into account, the previously anticipated trends are restored: the UVLF predicted by the ST model underpredicts the observational data, the DP1 model provides a modest improvement, and the DP2 model yields predictions in good agreement with the observational data.

At $z=9$, the observational data exhibit greater scatter and span a broader range of star formation efficiencies. Consequently, the ST model demonstrates reasonable agreement with at least a subset of the observed UV luminosity function (UVLF) data under conditions of high star formation efficiency. In contrast, the DP1 and DP2 models yield more accurate predictions for the observational data at intermediate star formation efficiencies, specifically at $0.1\lesssim f_\star \lesssim 0.5$ and $0.05\lesssim f_\star \lesssim 0.25$, respectively. The predictions incorporating $f_\star(M_h, z)$ follow a pattern similar to the two previous cases: the ST model systematically underpredicts the UVLF, especially at high luminosities, DP1 offers modest improvement, and DP2 shows good agreement with the observations.

At redshift $z=10$, the model predictions begin to show modest deviations. The UVLF based on the ST model aligns reasonably well with the data for $f_{\star}\simeq 0.25$, whereas the DP1 and DP2 models are more consistent with the observations for $f_{\star}\simeq 0.1$ and $f_{\star}\lesssim 0.05$, respectively. Although this trend differs somewhat from those observed at lower redshifts, it continues to underscore that high star formation efficiencies are not required when adopting more realistic halo mass functions. This is due to the fact that more realistic halo physics reduces the need for extreme astrophysical assumptions. In this particular case, the behavior of the parameter $f_\star(M_h, z)$ exhibits subtle differences. Specifically, the UVLF derived from the ST model demonstrates reduced deviation from the observational data compared to other redshift scenarios, although it still systematically underpredicts the observations. In contrast, the results obtained using the DP1 model, excluding $M_{\rm UV}\lesssim 20$, are in close agreement with observations. Conversely, the DP2 model yields a UVLF that significantly overpredicts the data at magnitude $M_{\rm UV}\lesssim 19$.

These findings reveal that the observational data themselves evolve only mildly in normalization from $z = 7$ to $9$, implying a surprisingly high abundance of luminous galaxies persisting into the earliest cosmic epochs. The fact that DP1 and DP2 can accommodate this with moderate $f_\star$ values, while ST cannot, strongly suggests that the tension between JWST and $\Lambda$CDM may not require new cosmology or exotic astrophysics, but rather a more accurate modeling of dark matter halo formation. The success of DP2 in particular highlights the critical role of non-spherical collapse dynamics and dissipative effects like dynamical friction in shaping the high-redshift galaxy population.

\begin{figure*}
\centering
\includegraphics[width=0.9\linewidth]{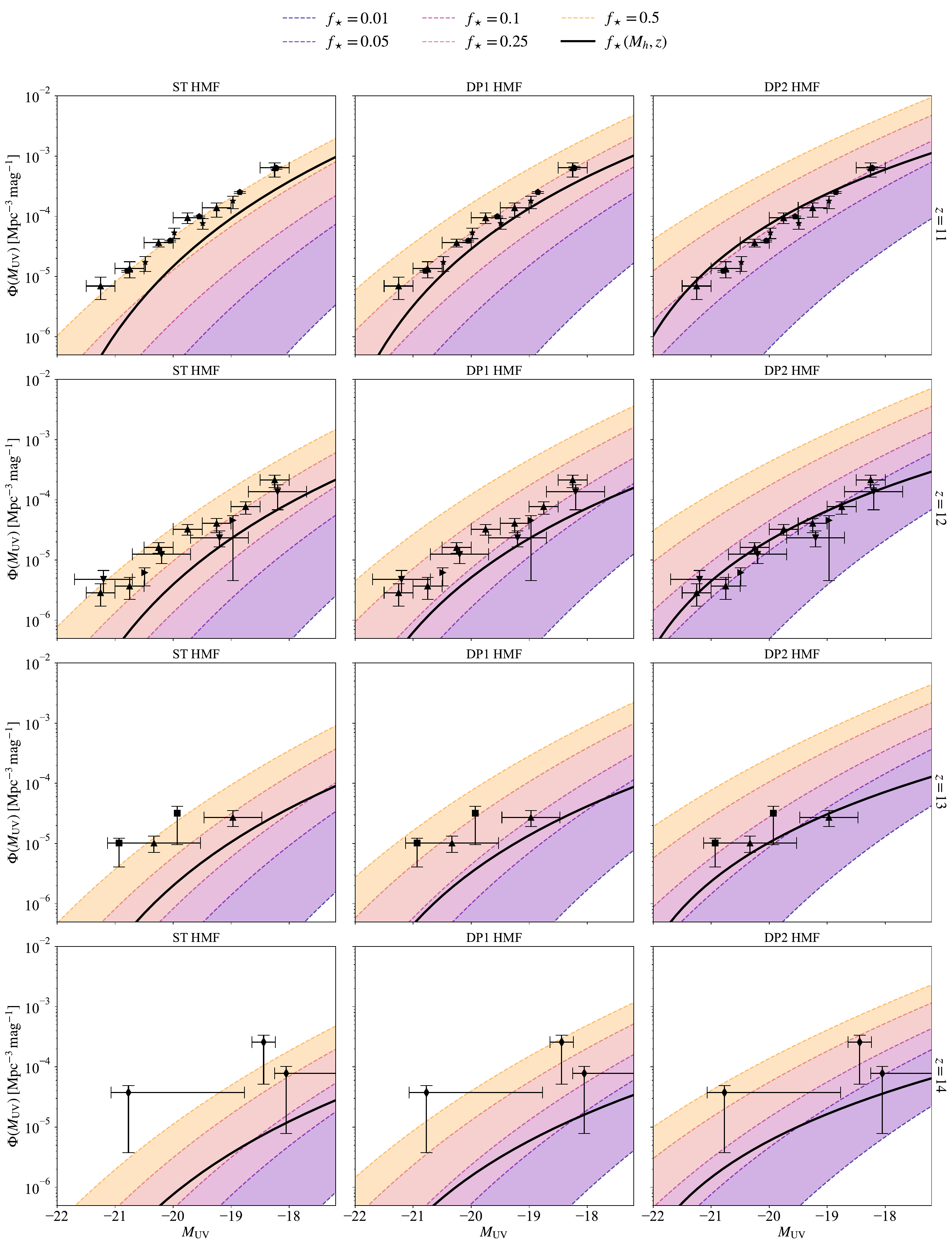}.
\caption{Similar to Fig.\,\ref{Fig2}, but for $z=11\mbox{-}14$. The dataset, displayed with error bars, presents recent observational constraints on the UV luminosity function at a common redshift, synthesized from a range of independent studies: filled squares correspond to the measurements reported in \citet{2015ApJ...803...34B, 2021AJ....162...47B, 2022ApJ...927...81B}; upward filled triangles denote data from \citet{2023MNRAS.520.4554D}; downward filled triangles represent results from \citet{2023ApJS..265....5H}; rightward filled triangles indicate findings from \citet{2024ApJ...965..169A}; filled stars correspond to \citet{2023ApJ...946L..13F}; filled pentagons signify the measurements presented in \citet{2024MNRAS.527.5004M}; and filled diamonds denote the constraints derived by \citet{2024ApJ...970...31R}.}
\label{Fig3}
\end{figure*}
In Fig.\,\ref{Fig3}, we have extended the analysis of Fig.\,\ref{Fig2} to even higher redshifts, $z = 11\mbox{-}14$, probing the epoch when the Universe was only $\sim$250-400 Myr old. As shown in Fig.\,\ref{Fig2}, the UVLF is plotted as a function of rest-frame UV magnitude. Theoretical predictions based on three HMFs ,ST, DP1 and DP2, are compared with recent high-redshift constraints from JWST. Results are shown for five constant star formation efficiencies ($f_\star=0.01,0.05,0.1,0.25,0.5$) as well as for a redshift- and halo mass-dependent efficiency, $f_\star(M_h, z)$. The left, middle, and right columns correspond to the ST, DP1, and DP2 models, respectively, while each row represents a fixed redshift from $z = 11$ (top) to $z = 14$ (bottom). We have also plotted the dataset with error bars reflecting recent observational constraints on the UV luminosity function at a common redshift, compiled from a variety of independent studies. Specifically, filled squares represent measurements reported in \citet{2015ApJ...803...34B, 2021AJ....162...47B, 2022ApJ...927...81B}; upward filled triangles correspond to data from \citet{2023MNRAS.520.4554D}; downward filled triangles indicate results from \citet{2023ApJS..265....5H}; rightward filled triangles denote findings from \citet{2024ApJ...965..169A}; filled stars refer to the constraints presented in \citet{2023ApJ...946L..13F}; filled pentagons signify measurements from \citet{2024MNRAS.527.5004M}; and filled diamonds represent the constraints derived by \citet{2024ApJ...970...31R}.

At $z=11$ , the UVLF predicted by the ST model, when coupled with a relatively high star formation efficiency of $f_{\star} = 0.5$, provides a suitable match to current observational data across the full range of UV magnitudes. This agreement, however, arises under the assumption of a highly efficient star formation scenario, which may be physically less plausible given constraints from lower redshift observations and theoretical models of early galaxy formation. In contrast, models incorporating more realistic prescriptions for star formation require significantly lower efficiencies to reproduce the observed UVLF at this epoch. Notably, the DP1 model with $f_{\star} = 0.25$ and the DP2 model with $f_{\star} = 0.1$ both yield good agreement with observational measurements, suggesting that more moderate star formation efficiencies are sufficient when physically motivated galaxy formation processes are included. Applying the factor $f_\star(M_h, z)$ to the calculations at this redshift yields the expected behavior. The UVLF predicted by the ST model still falls significantly below the observational data, whereas the DP1 and DP2 models reproduce the UVLF with markedly better agreement and smaller deviations from the JWST constraints.

At $z = 12$, the observational uncertainties associated with the UVLF are somewhat larger compared to those at lower redshifts. As a result, a broader range of model parameters, particularly the star formation efficiency can yield predictions that are consistent with the data. Within this context, the ST-based UVLF demonstrates good agreement with the observational constraints for star formation efficiencies in the range $0.25 \lesssim f_{\star} \lesssim 0.5$. However, the quality of this fit tends to be slightly better at the faint end of the UVLF than at the bright end, suggesting a possible tension in reproducing the observed number densities of the most luminous galaxies with high-efficiency models. A similar behavior is observed for the more physically motivated DP1 and DP2 models, though these require systematically lower star formation efficiencies to match the data. Specifically, the DP1 model aligns well with the observations for $0.1 \lesssim f_{\star} \lesssim 0.25$, while the DP2 model requires $0.05\lesssim f_{\star} \lesssim 0.1$ to maintain consistency. These trends underscore the sensitivity of UVLF predictions to assumptions about the underlying galaxy formation physics, particularly at the highest redshifts where observational constraints are still evolving. Furthermore, the ST and DP1 models exhibit nearly identical trends while considering $f_\star(M_h, z)$ and both underpredict the UVLF relative to the observational data, whereas the DP2 model yields significantly better agreement with the observations

At $z = 13$, the available UVLF measurements exhibit a preference for galaxy populations characterized by relatively high stellar mass densities, implying either enhanced stellar mass assembly in individual halos or a higher number density of faint, star-forming systems than predicted by minimal models. Despite this apparent demand for elevated stellar content, the overarching trend observed at lower redshifts persists: as the underlying halo model becomes increasingly physically motivated, the required star formation efficiency to reproduce the observed UVLF continues to decline. In other words, even at $z = 13$, more realistic models (e.g., DP1 and DP2) achieve satisfactory agreement with the data without invoking high values of $f_{\star}$, reinforcing the conclusion that early star formation is likely regulated by strong feedback and baryonic suppression mechanisms rather than proceeding at near-maximal efficiency. A qualitatively similar pattern is observed at $z = 14$, where preliminary UVLF constraints, though significantly sparser and subject to larger statistical and systematic uncertainties due to the extreme depth and cosmic variance limitations of current surveys, also appear consistent with models that assume low star formation efficiencies. At these high redshifts, adopting a halo mass- and redshift-dependent star formation efficiency does not fully alleviate the underprediction of the UVLF by the halo models relative to the observational data. However, as the HMF is refined from the ST prescription to the more physically motivated DP2 model, the discrepancy between theoretical predictions and observations can potentially be reduced.

These results further substantiate the interpretation that the apparent tension between JWST observations and the standard $\Lambda$CDM cosmological model does not inherently arise from fundamental cosmological discrepancies, but rather may be attributed to overly simplistic assumptions regarding the physics of halo formation. Notably, the DP2 model attains concordance with observational data without recourse to top-heavy initial mass functions (IMFs), non-standard dark matter candidates, or early dark energy scenarios, thereby highlighting the critical role of small-scale gravitational collapse processes in reconciling theory with observation.
\section{Conclusions}\label{sec:v}
In this work, we have confronted the emerging tension between early JWST observations of luminous high-redshift galaxies and the predictions of the standard $\Lambda$CDM cosmological framework by revisiting the foundational modeling of dark matter halo formation. Rather than invoking exotic cosmologies or extreme astrophysical assumptions, such as top-heavy IMFs, early dark energy, or non-standard dark matter, we have demonstrated that a more physically complete treatment of gravitational collapse within the $\Lambda$CDM paradigm can naturally reconcile theory with observation.

By employing refined HMFs, specifically the DP1 and DP2 models, that incorporate key physical effects absent in conventional approaches (e.g., ST), including angular momentum, and dynamical friction, we have shown a significant enhancement in the predicted abundance of massive halos at $z \gtrsim 7$. This enhancement is particularly pronounced in the high-mass tail $(M_h \gtrsim 10^9 \, M_\odot)$, which directly governs the bright end of the UVLF.

Our semi-empirical UVLF predictions, constructed via a physically motivated mapping from halo mass to star formation rate and UV luminosity, reveal that the DP2 model, augmented by dynamical friction, achieves suitable agreement with current JWST constraints across redshifts $z = 7$ to 14, even for moderate to low star formation efficiencies $(f_\star \sim 0.1\text{-}0.25)$. In stark contrast, the standard ST HMF requires unrealistically high efficiencies $(f_\star \gtrsim 0.5)$ to match the same data, thereby straining conventional models of early star formation and feedback.

Importantly, our analysis incorporates a physically motivated, redshift- and halo mass–dependent star formation efficiency, $f_\star(M_h, z)$, calibrated from empirical and theoretical studies of early galaxy formation. This prescription accounts for the expected suppression of star formation in low-mass halos due to photoheating and feedback, as well as potential evolution in efficiency with cosmic time. Our results indicate that even with this more realistic treatment, rather than assuming a constant, fine-tuned efficiency, the DP2 halo mass function continues to provide good agreement with JWST UVLF measurements across $z = 7\mbox{-}14$. In contrast, the ST model systematically underpredicts the bright-end UVLF even when using $f_\star(M_h, z)$, reinforcing our central conclusion that the resolution of the JWST overabundance tension lies primarily in improved modeling of halo collapse physics, not in extreme or ad hoc astrophysical assumptions.

Critically, our results indicate that the apparent overabundance of luminous galaxies in the epoch of reionization and cosmic dawn does not necessarily signal a breakdown of $\Lambda$CDM, but rather reflects the limitations of simplified halo collapse models that neglect dissipative and non-spherical dynamics. The success of the DP2 framework underscores the pivotal role of small-scale gravitational physics, particularly the lowering of effective collapse thresholds for rare, high-$\sigma$ peaks, in enabling earlier and more abundant formation of massive halos.

These findings advocate for a paradigm shift in theoretical galaxy formation: future models must integrate physically motivated, mass- and redshift-dependent collapse criteria to accurately capture the conditions of structure formation at the highest redshifts. Thus, our work can potentially resolve a key observational tension not through new physics beyond $\Lambda$CDM, but through a more faithful representation of the complex collapse dynamics that seed the first galaxies, a conclusion with profound implications for interpreting JWST data, modeling reionization, and forecasting next-generation high-redshift surveys.


\bigskip
\bibliography{draft_ml}
\end{document}